# Bell's theorem, entanglement, and common sense

Raoul Nakhmanson

*Nakhmanson@t-online.de*

A simple analogy to understand the quantum-mechanical entanglement within local-realistic world is presented.

The majority of professional physicists working in the "main stream" declare that quantum physics is a very strange domain which is not like anything we find in common life, and that it is unimaginable for our consciousness. They broadcast such a statement in professional and popular literature that inspire frustration and feeling of inferiority of their readers. Up to now used analogies with Carroll's personages are directed to show that these analogies don't work in accordance with common sense. The recent example devoted to *entanglement* was in [1].

Of course it would be better to find such an analogy which works. Then it can be used for education and popularization, and perhaps for physics itself.

Let's have a brief look at the history of the "main stream" meaning. The key words are *quantum mechanics* (*QM*), *reality, uncertainty, nonlocality, entanglement, interaction-free measurement*. They are interpreted as follows:

*QM* is a theory having its mathematical formalism which allows to describe the results of observations and experiments performed on separate microobjects (molecules, atoms, elementary particles, ...) and their ensembles. Some predictions of *QM* (eigenwerts, parameters) are very precise, others (results of interactions allowing alternative outcomes) are only probabilistic.

*Reality* is a physical world which stays behind the QM-formalism. The question on its immanent existence is declared by "main stream" as non-scientific. "No elementary phenomenon is a phenomenon until it is a registered (observed) phenomenon"[2], i.e. we have to think only about so-called "observables".

*Uncertainty* is principal impossibility of *QM* to forecast a result of an interaction e.g. a measurement having several possible outcomes. Only possibilities of corresponding outcomes can be calculated. This term is used also if two "non-commutable observables" (e.g. co-ordinate and impulse) want to be measured simultaneously. Such an *uncertainty* is a very special attribute of *QM* as compared with the determinism of classical physics.

*Nonlocality* is a possibility of an instant interaction between far parted objects.

*Entanglement* is such a correlation between once interacting and later departed particles which exceeds the possibility of local-realistic model with hidden parameters. The wave function of an entangled state can not be written as a product of individual wave functions.

*Interaction-free measurement* is a possibility to have some information about an object (its existence in some region of space, its form, ...) without any direct interaction between the object and a measurement apparatus.

The base of "main stream" philosophy was founded by Bohr. Planck, Einstein, de Broglie, Schrödinger, and others criticize the agnosticism of the "Copenhagen" interpretation of QM. Einstein, Podolsky and Rosen [3] tried to show the incompleteness of QM using the example with two particles (EPR Gedankenexperiment). De Broglie, Schrödinger and others tried to develop realistic models of quantum phenomena. An important work here was made by Bohm presenting in 1952 a consequent hidden-parameter model [4].

However the essential parameter of Bohm's model - quantum potential - was nonlocal for entangled states. Is nonlocality in any hidden-parameter model? In 1964 Bell apparently found an experimental possibility to prove it [5]. He showed that in EPR's situation the QM predicts a behavior correlation between far parted particles which are stronger than a correlation predicted by an arbitrary local-realistic model with hidden parameters. The following experiments [6,7] confirm the QM prediction. As it seems, trying to interpret the experiments one has a dilemma: either to reject any reality except "observables" like Bohr or to accept nonlocality like Bohm. Both alternatives are of course non-compatible with common sense.

So, must we follow the "main stream" and reject common sense? Must we use analogies and models which don't work in accordance with common sense? The answer is *not*, at least *not yet*, *not all*, and *not for education*.

Let us start with an analogy of the EPR-experiment. Supposing there are twins, Ralf and Rolf, living in Frankfurt and working for Lufthansa as pilots. For their employer (but not for their families!) they are "indistinguishable particles". The twins always try to dress alike, they believe that this brings them happiness. Because they are often in different countries, they agree on an order of sartorial priority: cold before warmth and rain before dry spell.

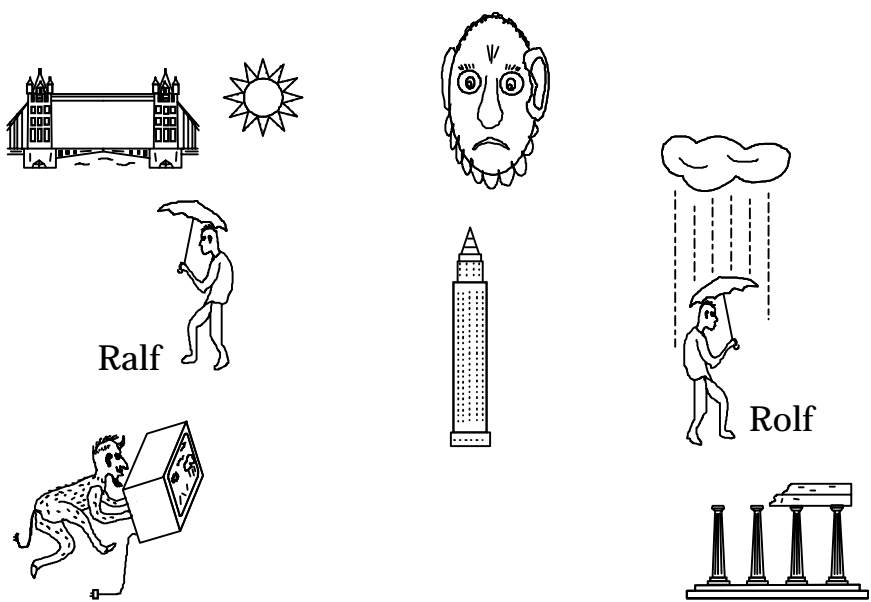

Fig. 1. Einstein-Podolsky-Rosen (EPR) experiment and the apparent
non-local interaction of entangled particles.

Nakhmanson 2

Once Ralf flies to England and Rolf flies to Greece (Fig.1). God who is observing the twins, sees that they carry clothes and shoes having the same sizes. He is not surprised: Ralf and Rolf are twins and have the same mechanical (hidden) parameters. God sees also that their sets of clothes and shoes are identical in fashion and color and very simple: each of them has only one variant for warm weather and one variant for cold weather. He does not wonder: such parameters can be stipulated in Frankfurt.

Besides that God sees a very striking correlation: the twins dress alike independent of the weather! It is natural if the weather in England and in Greece is the same. But why it is so when the weather is not the same? Why if it is cold in England, not only Ralf but also Rolf wears an overcoat in spite of the warm weather and if it is raining in Greece, not only Rolf but also Ralf hides beneath an umbrella, regardless of the sun in England (Fig.1)? "What is the matter?" - thinks God, - "I control experimental conditions, namely, today's weather in England and in Greece, and I know the twins' financial status, telephoning is too expensive for them. So it seems there is a non-local interaction (*entanglement*) between the twins. I am sure it is a new escapade of the devil!"

God's conclusion was only half true. In his heavenly chariot he fell behind the technical progress of the 20th century. He was right suspecting the devil. But up to now the devil does not realize non-local interaction. Instead, he had endowed humans with the ability *to forecast* future and also had invented radio, television, power computer for meteorology, and communication satellites for support. Therefore the twins using their own experience and radio or TV reports every evening can *forecast* the weather for tomorrow in England and Greece and dress in accordance with their agreement.

The difference between the twins' situation and the EPR-Bohm-Bell's analyse is the ability of living "particles" to forecast future. The living objects have a non-mechanical hidden parameter, namely, a *consciousness* !.

The proof of Bell's theorem is based on the next assertion: if $P_a$ is a probability of result *a* measured on the particle *1* in the point *A* having a condition (e.g. angle of analyzer) $\alpha$, and $P_b$ is a probability of result *b* measured on the particle *2* in the distant point *B* having a condition $\beta$, then $\beta$ has no influence on the $P_a$, and vice versa. Herein Bell and others saw the indispensable requirement of local realism and "separability". Mathematically it can be written as

$$P_{ab}(\lambda_{1i},\lambda_{2i},\alpha,\beta) = P_a(\lambda_{1i},\alpha) \times P_b(\lambda_{2i},\beta) \quad \text{(Bell)} , \quad (1)$$

where $P_{ab}$ is the probability of the join result *ab*, and $\lambda_{1i}$ and $\lambda_{2i}$ are hidden parameters of particles *1* and *2* in an arbitrary local-realistic theory. Under the influence of Bell's theorem and the experiments following it and showing, that for entangled particles the condition (1) is no longer valid, some "realists" reject locality. In this case an instantaneous action at a distance is possible, and one can write

$$P_{ab}(\lambda_{1i},\lambda_{2i},\alpha,\beta) = P_a(\lambda_{1i},\alpha,\beta) \times P_b(\lambda_{2i},\beta,\alpha) \quad \text{(non-locality)} . \quad (2)$$

In principle such a relation permits a description of any correlation between *a* and *b*, particularly predicted by QM and observed in experiments. But for *living particles* in the frame of local realism the condition (1) is not indispensable. Instead, one can write

$$P_{ab}(\lambda_{1i},\lambda_{2i},\alpha,\beta) = P_a(\lambda_{1i},\alpha,\beta') \times P_b(\lambda_{2i},\beta,\alpha') \quad \text{(forecast)}, \quad (3)$$

where $\alpha'$ and $\beta'$ are the conditions of measurements in points *A* and *B*, respectively, as they can be forecast by particles at the moment of their parting. If the forecast is good enough, i.e., $\alpha' \approx \alpha$ and $\beta' \approx \beta$, then (3) practically coincides with (2) and has all its advantages plus locality.

Persisting in "main stream" one can successfully use the analogy of the twins in an education process and conclude e.g. with "If a baby, having more experience with its parents than with "inanimate" matter, could make experiments, the behavior of microparticles would appear to it to be very natural".

Is the above analogy more than an analogy? Can "inanimate" matter really own some "consciousness" and forecast the future? Some philosophers and physicists supposed it [8]. There is a development of this idea to the level of a consequent hypothesis which can be tested experimentally [9-11]. For example, in EPR-type experiments one can quickly change an analyzer direction. Periodical switching [7] must be avoided because it is predictable, random switching is preferable [9-12]. The corresponding random generator must be "good" enough to exclude forecasting by the particles, but this issue is not trivial. For example Weihs *et al*. [12] wrote:

"Selection of an analyzer direction has to be completely unpredictable, which necessitates a physical random number generator. A pseudo-random-number generator cannot be used, since its state at any time is predetermined" (p. 5039). As a "physical random number generator" the authors used a "light-emitting diode illuminating a beam splitter whose outputs are monitored by photomultipliers" (p. 5041).

Such a point of view must be commented. It is a right prerequisite that the generator must be unpredictable for particles, but referring to above "physical" generator it is not clear: If particles have "consciousness" such a generator can also be a "pseudo" one. Perhaps a good "human" pseudo-random generator is preferable because it belongs to another civilization.